\documentclass[12pt]{article}

\usepackage{booktabs}
\usepackage{siunitx}
\usepackage{amsmath}

\usepackage{algorithm}
\usepackage{algpseudocode}

\usepackage{etoolbox}
\usepackage{bm}

\usepackage{arydshln}

\makeatletter
\def\mathcolor#1#{\@mathcolor{#1}}
\def\@mathcolor#1#2#3{%
	\protect\leavevmode
	\begingroup
	\color#1{#2}#3%
	\endgroup
}
\makeatother

\usepackage{url}
\usepackage{setspace}
\usepackage[table,xcdraw]{xcolor}

\usepackage{graphicx}
\usepackage{multirow}

\graphicspath{ {./figures/} }

\usepackage{caption}
\usepackage{subcaption}

\usepackage{amsmath} 
\usepackage{amsfonts}
\usepackage{nicefrac}

\usepackage[a4paper, total={6in, 8in}]{geometry}

\usepackage[symbol]{footmisc}

\usepackage[
colorlinks=true,
linkcolor=blue,
filecolor=blue,
urlcolor=blue,
citecolor=blue,
pagebackref=true
]{hyperref}
\makeatletter
\renewcommand*{\backref}[1]{}
\renewcommand*{\backrefalt}[4]{%
	\ifcase #1
	(Not cited.)%
	\or
	\textsuperscript{#2}%
	\else
	\textsuperscript{#2}%
	\fi}
\makeatother

\usepackage{xspace}   

\begin{document}
	\begin{center}	
		\textbf{\large {Image-Domain CT Metal Artifact Reduction Using an Endpoint-Conditioned Diffusion Bridge}} \\
		Hang Yu$ ^1 $, Mojtaba Safari$ ^2 $, Shansong Wang$ ^2 $, Mingzhe Hu$ ^2 $, and Xiaofeng Yang$ ^{2, \ddagger} $ \\
    \end{center}
\begin{flushleft}
$ ^1 $Department of Physics and Astronomy, University of North Carolina at Chapel Hill, Chapel Hill, NC, United States,\\	
$ ^2 $Department of Radiation and Cellular Oncology, The University of Chicago, Chicago, IL, United States,\\

\vspace{1 cm}

$ ^\ddagger  $Corresponding Author: \url{xfyang@uchicago.edu}
\end{flushleft}

\newpage

\begin{abstract}

\noindent\textbf{Background:}
Metal implants can generate severe streaking and shading artifacts
in computed tomography (CT), obscuring nearby anatomy and altering
local Hounsfield-unit (HU) values. Conventional diffusion models
typically map images toward a Gaussian distribution and initialize
reverse reconstruction from Gaussian noise, requiring the target
image to be recovered through iterative denoising from an
uninformative starting state. For CT restoration, this formulation
does not directly exploit anatomical information already present
in the metal-artifact image.

\noindent\textbf{Purpose:}
To adapt a diffusion bridge framework for image-domain CT metal
artifact reduction (MAR) by directly connecting paired
metal-artifact and metal-free CT images. The proposed RDBM-MAR
model initializes reverse reconstruction from the metal-artifact
CT image rather than Gaussian noise, aiming to retain available
anatomical information and learn a direct restoration trajectory
toward the metal-free target.

\noindent\textbf{Methods:}
RDBM-MAR constructs an endpoint-conditioned diffusion bridge
between paired metal-free and metal-artifact CT images. During
training, intermediate bridge states are generated between the
two endpoints, and a conditional U-Net receives the bridge state,
metal-artifact image, and timestep to estimate the metal-free CT
image through an image-domain correction. The forward bridge
used 100 timesteps, while inference used 10 deterministic reverse
steps initialized from the metal-artifact image. Training,
validation, and quantitative testing used 14,000 paired
hybrid-simulated CT slices from the AAPM CT Metal Artifact
Reduction Grand Challenge. These slices were derived from the
DeepLesion dataset and the UCLH Stroke EIT Dataset, with virtual
metal objects and artifacts generated using CatSim in the
open-source XCIST toolkit. The data were divided into 12,045
training, 633 validation, and 1,322 held-out test slices.
One case from the 29-case AAPM CT-MAR final scoring dataset
was used for qualitative evaluation. RDBM-MAR was compared with
CycleGAN, Pix2Pix, Res-SRDiff, I2I-Mamba, and I2SB using PSNR,
SSIM, MAE, and LPIPS. Statistical analysis used the Friedman
test followed by paired Wilcoxon signed-rank tests with Holm
correction.

\noindent\textbf{Results:}
RDBM-MAR achieved the highest PSNR ($43.67 \pm 4.56$ dB) and
SSIM ($0.9856 \pm 0.0138$) and the lowest MAE
($14.74 \pm 8.68$ HU) among the evaluated methods.
Significant between-method differences were observed for all
four metrics ($p < 0.001$). Pairwise Wilcoxon signed-rank
tests with Holm correction showed significantly better PSNR,
SSIM, and MAE for RDBM-MAR than for all five comparison methods
(adjusted $p < 0.001$). Res-SRDiff achieved the lowest LPIPS.
Difference maps and HU line-profile analysis showed reduced
artifact-related intensity errors with preservation of local
anatomical structures. The AAPM final scoring case also showed
qualitative reduction of visible metal-associated streaking.

\noindent\textbf{Conclusions:}
RDBM-MAR provides a direct restoration trajectory from the
metal-artifact CT image toward the metal-free target while
retaining anatomical information in the input. The proposed
formulation achieved strong structural fidelity and HU accuracy
on the held-out paired test set using 10 reverse sampling steps,
supporting diffusion bridge modeling for image-domain CT MAR.

\end{abstract}


\section{Introduction}
\label{sec:intro}

Computed tomography (CT) is a cornerstone of clinical imaging, providing detailed visualization of anatomical structures together with quantitative tissue information through the Hounsfield unit (HU) scale. Metal implants, however, violate the assumptions underlying standard reconstruction: beam hardening, photon starvation, scatter, and the resulting inconsistencies in the projection data generate bright and dark streaks that radiate from the metal~\cite{gjesteby2016mar}. These artifacts distort local HU values, obscure the anatomy immediately adjacent to the implant, and propagate into downstream tasks such as segmentation, anatomical localization, and quantitative image analysis. Metal artifact reduction (MAR) has therefore been studied extensively, yet suppressing artifacts while simultaneously preserving local anatomical structure and quantitative CT intensity remains an open problem~\cite{gjesteby2016mar,kleber2024review}.

Approaches to MAR fall broadly into projection-domain and image-domain families. Projection-domain methods, including linear interpolation~\cite{kalender1987li}, normalized MAR~\cite{meyer2010nmar}, and their learning-based successors~\cite{zhang2018cnnmar}, correct the inconsistent measurements at their physical origin and can therefore be highly effective, but they require access to raw projection data together with the scanner geometry, bowtie filter, and calibration parameters used for reconstruction. Such data are proprietary and are rarely exported in clinical practice, where the archived product is the reconstructed image; models developed for one scanner geometry additionally transfer poorly to another. Errors introduced during sinogram completion also propagate along entire ray paths and can manifest as secondary artifacts remote from the implant~\cite{lin2019dudonet}, whereas corrections applied in the image domain remain spatially local. We therefore adopt an image-domain formulation, which is directly applicable to retrospective and clinical data for which only reconstructed images are available.

Within the image domain, a network learns the mapping from artifact-corrupted CT images to their metal-free counterparts from paired training data. Conditional image-to-image translation models~\cite{isola2017pix2pix} achieve strong performance on paired restoration tasks and have been applied directly to MAR~\cite{wang2018conditional}, suppressing prominent streaks and restoring the appearance of corrupted regions. Their adversarial objectives, however, are not intensity-preserving by construction, and reported failure modes include local HU shifts, residual streaking, and over-smoothing of fine anatomical detail~\cite{wang2018conditional,lin2019dudonet}. This limitation reflects a broader tension between perceptual quality and distortion fidelity~\cite{blau2018tradeoff}, and it is consequential in MAR, where the corrected image is frequently used for quantitative measurement.

Diffusion models address part of this problem by modeling restoration as a gradual stochastic transformation between image distributions~\cite{ho2020ddpm,nichol2021improved}, but in the conventional formulation sampling begins from Gaussian noise and the degraded observation enters only as conditioning information, even though that observation already carries most of the anatomical and intensity content of the target. A more efficient alternative is to let the residual between the paired images drive the process. We previously showed that shifting the forward process along the residual between low- and high-resolution MRI aligns the two distributions directly and permits high-fidelity reconstruction in as few as four sampling steps, at a fraction of the cost of conditional diffusion~\cite{safari2025ressrdiff}. Related formulations construct an explicit stochastic bridge between the paired endpoints~\cite{liu2023i2sb,wang2026rdbm}, and allow the residual, where available, to modulate where noise is injected and removed, so that perturbation concentrates on degraded regions while intact anatomy is left largely undisturbed~\cite{wang2026rdbm}.

Metal artifacts differ from the natural-image degradations considered in previous diffusion bridge studies. They are signed and non-local, correlated along ray paths rather than confined to a compact region, and the restoration
target is a quantitative HU value rather than a plausible visual appearance. For CT metal artifact reduction, we therefore use the uniform-modulation Ornstein--Uhlenbeck bridge formulation. The paired clean-to-artifact difference
remains in the bridge mean, while the Gaussian perturbation is spatially uniform. This provides a direct stochastic transition between the paired metal-free and metal-artifact endpoints without residual-dependent noise
modulation.

In the proposed method, the metal-free CT image $x_0$ and its paired metal-artifact image $\mu$ are defined as the two endpoints of the diffusion bridge, following the notation of Wang \emph{et al.}~\cite{wang2026rdbm}, in which $\mu$ denotes the degraded observation rather than the linear attenuation coefficient. During training, intermediate bridge states are sampled between the endpoints, and a conditional U-Net~\cite{ronneberger2015unet} receives the current bridge state, the metal-artifact image, and the timestep, and estimates the corresponding metal-free CT image. At inference, the reverse process is initialized directly from the observed metal-artifact image and recovers the clean image along a shortened sampling trajectory. Because the network predicts an image-domain correction relative to the degraded input rather than synthesizing a complete CT image from Gaussian noise, the restoration proceeds directly from the observed metal-artifact image toward the metal-free target while retaining the anatomy already present in the input.

Our main contributions are as follows:

\begin{itemize}
  \item We adapt an endpoint-conditioned Ornstein-Uhlenbeck diffusion bridge to image-domain CT metal artifact reduction, in which the artifact-corrupted scan serves as a fixed endpoint of the stochastic process and the network predicts an image-domain correction relative to the degraded input, preserving the anatomy and HU content already present in the scan. The contribution is the MAR-specific instantiation and systematic evaluation, rather than the bridge formulation itself.

\item We use the uniform-modulation bridge formulation, in which the paired clean-to-artifact difference contributes to the bridge mean while the Gaussian perturbation remains spatially uniform, avoiding residual-dependent stochastic
modulation while retaining direct conditioning on the paired endpoints.

\item We benchmark the proposed method against representative methods from four distinct restoration families including adversarial (Pix2Pix~\cite{isola2017pix2pix}, CycleGAN~\cite{zhu2017cyclegan}), diffusion-based (Res-SRDiff~\cite{safari2025ressrdiff}), bridge-based (I$^2$SB~\cite{liu2023i2sb}), and state-space (I2I-Mamba~\cite{atli2026i2imamba}), all trained and evaluated under identical data splits and preprocessing on a common held-out CT-MAR test set.

  \item We evaluate both fidelity and perceptual quality using peak signal-to-noise ratio (PSNR), structural similarity index measure (SSIM)~\cite{wang2004ssim}, mean absolute error (MAE), and learned perceptual image patch similarity (LPIPS)~\cite{zhang2018lpips}, and report slice-level metric distributions with paired statistical comparisons rather than cohort means alone, so that differences between methods are assessed for consistency as well as magnitude.

  \item We separate artifact suppression from quantitative intensity recovery through localized difference maps and HU line-profile analysis, addressing a distinction that aggregate image-quality metrics do not capture.

  \item We assess generalization beyond the simulated training distribution on clinical CT data from the AAPM CT-MAR benchmark~\cite{haneda2025aapm,peters2025hybrid}.

\end{itemize}

\section{Materials and methods}
\subsection{Data Preparation}\label{subsec:data_preparation}

The proposed model was developed and evaluated on the paired image-domain CT data released by the AAPM CT Metal Artifact Reduction (CT-MAR) grand challenge~\cite{haneda2025aapm,aapmbenchmark}. The challenge database was constructed with a hybrid simulation framework~\cite{peters2025hybrid} that inserts virtual metal objects into clinical CT images drawn from DeepLesion~\cite{yan2018deeplesion} and the UCLH Stroke EIT dataset~\cite{goren2017uclh}, covering head and body anatomy. Artifact-corrupted projections and their reconstructions were generated with the CatSim simulator in the open-source XCIST toolkit~\cite{wu2022xcist}, which models quantum noise, beam hardening, and scatter. Each sample therefore comprises a metal-artifact CT image and its spatially registered metal-free reference, enabling supervised training in the image domain.

All images were processed using the same preprocessing procedure. CT values were clipped to the range of $[-1000,2500]$~HU to standardize the intensity range and retain diagnostically relevant tissue information, followed by linear normalization to $[0,1]$. The identical normalization procedure was applied to both the metal-artifact input images and their corresponding metal-free references. No spatial resizing was performed, and all images were retained at their native $512\times512$ resolution.

A total of 12,045 paired samples were used for training, while 633 paired samples were allocated for validation. A separate held-out test set containing 1,322 paired samples was reserved exclusively for final model evaluation and was not used during model training or model selection. The pairing between each metal-artifact image and its corresponding metal-free reference was strictly maintained throughout data partitioning and evaluation.


For qualitative assessment, one clinical case from the AAPM CT-MAR scoring set was additionally evaluated~\cite{haneda2025aapm,peters2025hybrid}. This case contained a metal implant, had no metal-free ground truth, and was not used during training. Metal regions were delineated from the unclipped HU image by thresholding at 3000~HU and are overlaid in red on all displayed images; the resulting mask was used for visualization only and was never supplied to the model during training or inference.

\begin{figure}[!t]
    \centering
    \includegraphics[width=1\linewidth]{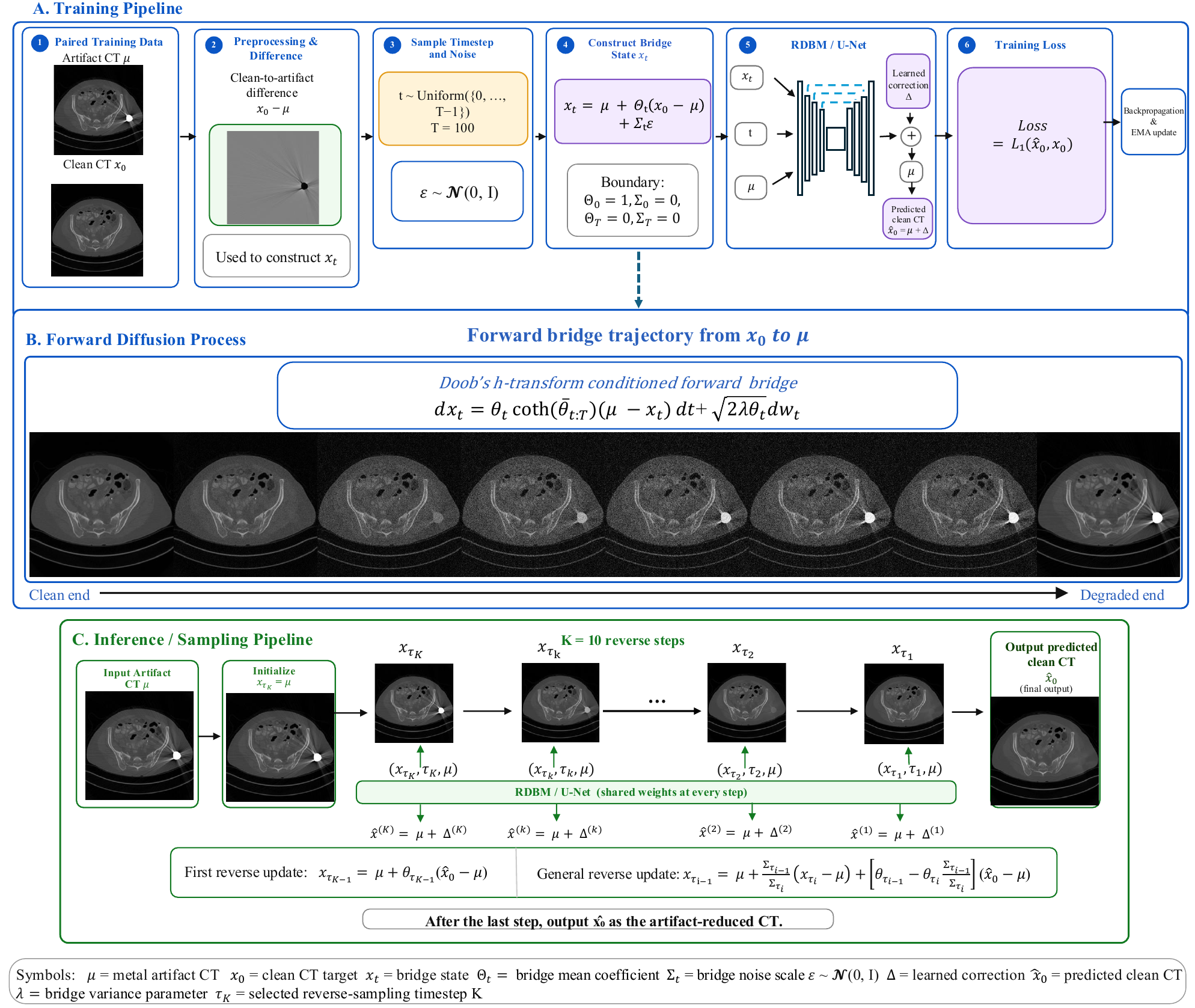}
    \caption{Overview of the proposed diffusion bridge framework for image-domain CT metal artifact reduction. (A) During training, paired metal-free CT images $x_0$ and metal-artifact CT images $\mu$ are used to construct intermediate bridge states, and a conditional U-Net predicts a correction to obtain the clean-image estimate $\hat{x}_0$. (B) The forward bridge connects $x_0$ to the degraded endpoint $\mu$ under the Doob's $h$-transform-conditioned process. (C) During inference, reverse sampling starts from the observed metal-artifact image and reconstructs the artifact-reduced CT image using $K=10$ reverse steps.}
    \label{fig:model_framework}
\end{figure}

\label{sec:ctmar_bridge}

\subsection{CT Metal Artifact Diffusion Bridge}
\label{sec:ctmar_bridge}

For each paired CT sample, the metal-free image is denoted by $x_0$ and the corresponding metal-artifact image by $\mu$. The forward process establishes a stochastic transition from the metal-free image to its degraded counterpart, whereas the reverse process reconstructs the metal-free image beginning from the observed metal-artifact CT.


We adopt the generalized Ornstein--Uhlenbeck bridge~\cite{yue2024goub}, in which Doob's $h$-transform conditions a mean-reverting OU process on the fixed terminal state $x_T=\mu$, eliminating the terminal variance and yielding a point-to-point mapping between the paired images. This formulation is the uniform-modulation case of the generalized bridge family described by Wang \emph{et al.}~\cite{wang2026rdbm}: the stochastic modulation term is held at unity, so the Gaussian perturbation is spatially uniform rather than residual-weighted. The resulting process is

\begin{equation}
    d x_t = \theta_t \coth\left(\bar{\theta}_{t:T}\right) (\mu-x_t)\,dt + \sqrt{2\lambda\theta_t}\,d\omega_t, 
\end{equation}
where $\theta_t$ is the time-dependent drift coefficient, $\bar{\theta}_{t:T}=\int_t^T\theta_s\,ds$ is the accumulated drift over the remaining interval, $\lambda$ scales the bridge variance, and $\omega_t$ is a standard Wiener process.

The forward trajectory need not be simulated: an intermediate bridge state admits the closed form

\begin{equation}
\centering
x_t = \mu + \Theta_t(x_0-\mu) + \Sigma_t\epsilon, \qquad \epsilon\sim\mathcal{N}(0,I), 
\label{eq:bridge_state}
\end{equation}

with coefficients

\begin{equation}
\Theta_t = \frac{\sinh\!\left(\bar{\theta}_{t:T}\right)}
{\sinh\!\left(\bar{\theta}_{0:T}\right)},
\qquad
\Sigma_t^2 = 2\lambda\,
\frac{\sinh\!\left(\bar{\theta}_{0:t}\right)
\sinh\!\left(\bar{\theta}_{t:T}\right)}
{\sinh\!\left(\bar{\theta}_{0:T}\right)} .
\label{eq:bridge_coefficients}
\end{equation}

Here $\Theta_t$ weights the metal-free-to-artifact difference $x_0-\mu$ and $\Sigma_t$ sets the perturbation magnitude. The construction satisfies both endpoint conditions: at $t=0$, $\Theta_0=1$ and $\Sigma_0=0$, recovering $x_0$; at $t=T$, $\Theta_T=0$ and $\Sigma_T=0$, so the state collapses to $\mu$ with no residual noise. The drift schedule $\theta_t$ follows a cosine profile~\cite{nichol2021improved}, with $\lambda=10^{-4}$ and $T=100$ discretization steps.  A conditional U-Net receives the current bridge state $x_t$, the metal-artifact image $\mu$, and the corresponding time variable. The network predicts an image-domain correction $\Delta_\phi$, which is combined with the metal-artifact image to obtain the metal-free estimate,

\begin{equation}
\hat{x}_0 = \mu + \Delta_\phi(x_t,\mu,t).
\label{eq:clean_prediction}
\end{equation}

This residual parameterization confines the learning problem to the artifact-induced component of the image, so that anatomy already present in $\mu$ is carried through unaltered. The network is trained by minimizing an $\ell_1$ loss between $\hat{x}_0$ and the paired metal-free reference $x_0$, computed on the normalized $[0,1]$ intensity scale over all image pixels.

Inference uses a shortened trajectory $\{\tau_K,\dots,\tau_0\}$ of $K=10$ steps selected by uniform spacing from the training discretization, and is initialized directly from the observation,

\begin{equation}
x_{\tau_K}=\mu .
\label{eq:reverse_initialization}
\end{equation}

Because $\Sigma_T=0$, the general update below is undefined at the first step, which is therefore evaluated directly from Eq.~\eqref{eq:bridge_state} with the noise term omitted:

\begin{equation}
x_{\tau_{K-1}} = \mu + \Theta_{\tau_{K-1}}\!\left(\hat{x}_0^{(K)}-\mu\right).
\label{eq:first_reverse}
\end{equation}

Subsequent states follow the deterministic update

\begin{equation}
\begin{split}
x_{\tau_{i-1}} ={}& \mu
+ \frac{\Sigma_{\tau_{i-1}}}{\Sigma_{\tau_i}}\left(x_{\tau_i}-\mu\right) \\
&+ \left[\Theta_{\tau_{i-1}}
- \Theta_{\tau_i}\frac{\Sigma_{\tau_{i-1}}}{\Sigma_{\tau_i}}\right]
\left(\hat{x}_0^{(i)}-\mu\right).
\end{split}
\label{eq:reverse_update}
\end{equation}

No stochastic term is injected during reverse sampling, so reconstruction is deterministic given $\mu$ and reproducible across runs. The estimate at $\tau_0$ is returned as the metal-artifact-reduced image and mapped back to HU by inverting the normalization of Section~\ref{subsec:data_preparation}.

\begin{algorithm}[t]
\caption{Training of the CT metal artifact diffusion bridge}
\label{alg:training}
\begin{algorithmic}[1]
\Require Paired training set $\mathcal{D}=\{(x_0,\mu)\}$; network $\Delta_\phi$;
         bridge coefficients $\Theta_t,\Sigma_t$ from
         Eq.~\eqref{eq:bridge_coefficients}
\Repeat
  \State Sample a minibatch $(x_0,\mu)\sim\mathcal{D}$
  \State Sample $t \sim \mathcal{U}\{0,\ldots,T-1\}$
  \State Sample $\epsilon\sim\mathcal{N}(0,I)$
  \State $x_t \gets \mu+\Theta_t(x_0-\mu)+\Sigma_t\epsilon$
         \Comment{Eq.~\eqref{eq:bridge_state}}
  \State $\hat{x}_0 \gets \mu+\Delta_\phi(x_t,\mu,t)$
         \Comment{Eq.~\eqref{eq:clean_prediction}}
  \State $\phi \gets \phi-\eta\,\nabla_\phi\,
         \mathcal{L}\!\left(\hat{x}_0,x_0\right)$
\Until{convergence}
\State \Return $\phi$
\end{algorithmic}
\end{algorithm}

\begin{algorithm}[t]
\caption{Reverse sampling for CT metal artifact reduction}
\label{alg:sampling}
\begin{algorithmic}[1]
\Require Metal-artifact image $\mu$; trained network $\Delta_\phi$;
         step sequence $\tau_K>\cdots>\tau_0$; coefficients
         $\Theta_{\tau_i},\Sigma_{\tau_i}$
\State $x_{\tau_K}\gets\mu$
\For{$i=K,K-1,\dots,1$}
  \State $\hat{x}_0^{(i)}\gets\mu+\Delta_\phi(x_{\tau_i},\mu,\tau_i)$
  \If{$i=K$}
    \State $x_{\tau_{i-1}}\gets
           \mu+\Theta_{\tau_{i-1}}\!\left(\hat{x}_0^{(i)}-\mu\right)$
           \Comment{Eq.~\eqref{eq:first_reverse}; $\Sigma_{\tau_K}=0$}
  \Else
    \State Update $x_{\tau_{i-1}}$ by Eq.~\eqref{eq:reverse_update}
  \EndIf
\EndFor
\State \Return $x_{\tau_0}$
       \Comment{$\Theta_{\tau_0}=1,\ \Sigma_{\tau_0}=0
                \Rightarrow x_{\tau_0}=\hat{x}_0^{(1)}$}
\end{algorithmic}
\end{algorithm}

\subsection{Implementation Details}

The forward bridge was discretized into $T=100$ timesteps with bridge variance $\lambda=10^{-4}$. At each training iteration, a timestep $t$ was drawn uniformly from
$\{0,\ldots,T-1\}$ and Gaussian noise $\epsilon \sim \mathcal{N}(0,I)$ was used to construct the corresponding intermediate bridge state via Eq.~\eqref{eq:bridge_state}. Because a single conditional network is shared across all timesteps, this sampling exposes the same weights to the full range of intermediate corruption levels, and the same trained model is applied at every reverse stage.

The network was trained using direct clean-image supervision. Specifically,
the final prediction was parameterized as an additive image-domain correction
relative to the observed metal-artifact image,

\begin{equation}
\hat{x}_0 = \mu+\Delta_\phi(x_t,\mu,t),
\end{equation}

and optimized using an $L_1$ reconstruction objective,

\begin{equation}
    \mathcal{L}_{1} = \left\| \hat{x}_0-x_0 \right\|_1. 
\end{equation}


The model was optimized using Adam~\cite{kingma2014adam} with a learning rate of $8\times10^{-5}$ and exponential decay coefficients $(\beta_1,\beta_2)=(0.9,0.99)$. Training was performed for 120,000 iterations using a batch size of 10. Gradient clipping with a maximum norm of 1.0 was applied during optimization. An exponential moving average of the model parameters was maintained with a decay factor of 0.995 and used for validation and inference.

At inference the schedule was reduced to $K=10$ reverse steps, selected by uniform spacing from the training discretization. Reconstruction was initialized from the observed metal-artifact image and no stochastic term was injected, so inference is deterministic for a fixed input and parameter set. All slices were processed at their native $512\times512$ resolution, and outputs were mapped back to HU by inverting the normalization described in Section~\ref{subsec:data_preparation} before metric computation.

\subsection{Comparison Methods}

The proposed diffusion bridge model was evaluated against five representative image restoration approaches: Pix2Pix~\cite{isola2017pix2pix}, CycleGAN~\cite{zhu2017cyclegan}, Res-SRDiff~\cite{safari2025ressrdiff}, I2I-Mamba~\cite{atli2026i2imamba}, and I$^2$SB~\cite{liu2023i2sb}. These methods were selected to represent different image-to-image restoration paradigms and to provide comparisons with both conventional and more recent deep-learning approaches.

Pix2Pix provides a supervised conditional generative adversarial network baseline that directly learns a mapping between paired source and target images. CycleGAN represents an unpaired adversarial translation framework and therefore does not require one-to-one correspondence between source and target images during optimization. Res-SRDiff provides a residual-shifting diffusion-based restoration approach. I2I-Mamba uses selective state-space modeling for image-to-image transformation, while I$^2$SB formulates paired image translation as a Schr\"odinger bridge between the source and target distributions. Together, these models cover adversarial, diffusion-based, state-space-based, and bridge-based image restoration strategies.

For all comparison methods, the metal-artifact CT images were used as the source images and the corresponding metal-free CT images as the restoration targets. The same training, validation, and held-out test partitions were used across all methods, together with identical HU clipping, intensity normalization, and image resolution. For supervised approaches, the original artifact--metal-free pairing was preserved during training. CycleGAN used the same source and target training sets but did not use their paired correspondence, consistent with its original unpaired training formulation.

Each comparison model was adapted only as required to accommodate the single-channel $512\times512$ CT data and the artifact-to-metal-free restoration task while retaining its original learning formulation as closely as possible. Model selection was performed using the corresponding validation set, and final quantitative evaluation was conducted exclusively on the same 1,322 held-out test slices. The identical metal-free reference images and evaluation procedure were used for all methods to ensure direct slice-level comparison.

\subsection{Quantitative and Statistical Analysis}

Quantitative reconstruction performance was evaluated using PSNR, SSIM, MAE, and learned LPIPS. Higher PSNR and SSIM values indicate better performance, whereas lower MAE and LPIPS values indicate better performance.

Each metric was calculated independently for every held-out test slice using the corresponding metal-free CT image as the reference. PSNR and SSIM were computed using the same normalized intensity representation for all evaluated methods. For MAE calculation, reconstructed and reference images were mapped back to the original HU scale, and MAE was therefore reported in HU. For LPIPS evaluation, the single-channel CT images were replicated across three channels and mapped to the input range required by the pretrained LPIPS network. The same evaluation pipeline was applied to all reconstruction methods. Quantitative results were summarized as mean $\pm$ standard deviation, and slice-level metric distributions were visualized using boxplots.

Statistical comparisons were performed using nonparametric paired analyses. 
For each image-quality metric, differences among the six reconstruction
methods were first assessed using the Friedman test for repeated measures,
with each held-out test slice treated as a matched observation across methods.
The metal-artifact input was not included in the Friedman analysis.
Kendall's coefficient of concordance ($W$) was calculated to quantify the
effect size of the overall between-model differences.

Following a significant Friedman test, pairwise comparisons between the
proposed method and each of the five comparison models were performed using
the Wilcoxon signed-rank test on the paired slice-level metric values. Holm
correction was applied separately within each image-quality metric to account
for multiple comparisons. An adjusted $p<0.05$ was considered statistically
significant. Median paired differences were additionally reported for the
closest-performing comparison method to describe the magnitude and direction
of the observed differences.

\section{Results}

\subsection{Qualitative Comparison}

\begin{figure}[!tbh]
    \centering
    \includegraphics[width=\textwidth]{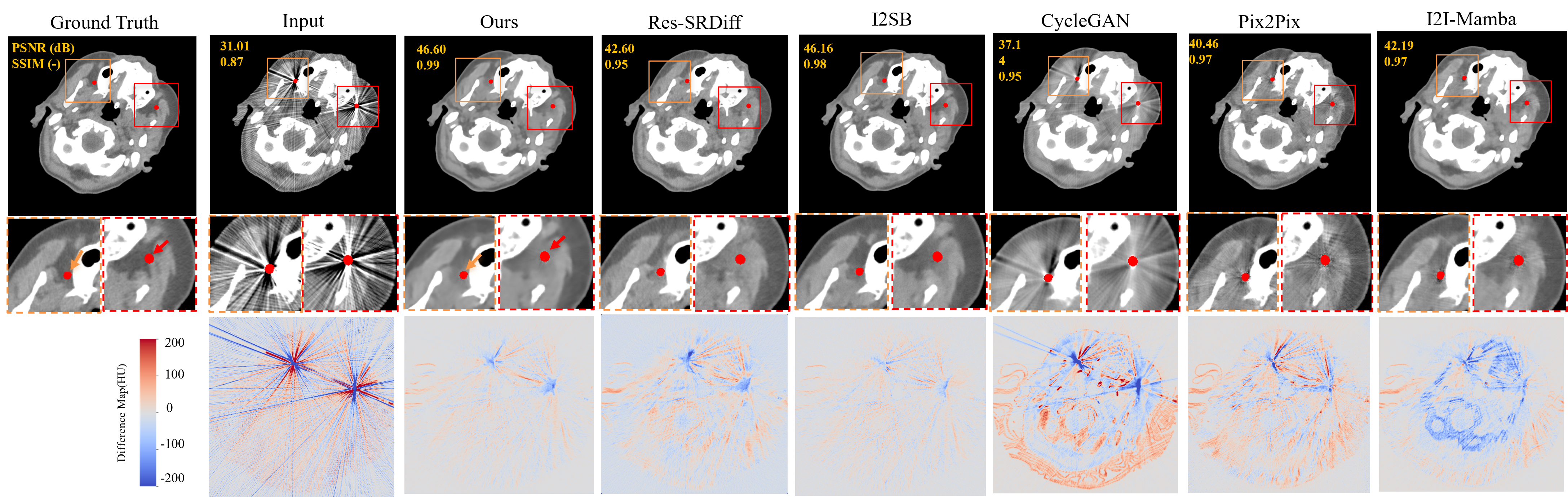}
    \caption{Qualitative comparison on representative head CT from the held-out test set. Reconstructed images and corresponding    difference maps relative to the metal-free reference are shown for the evaluated methods. Red regions indicate metallic implants. CT images are displayed using a window of $[-300,300]$ HU, and difference maps are reported in HU.}
    \label{fig:result1}
\end{figure}

\begin{figure}[!tbh]
    \centering
        \includegraphics[width=\textwidth]{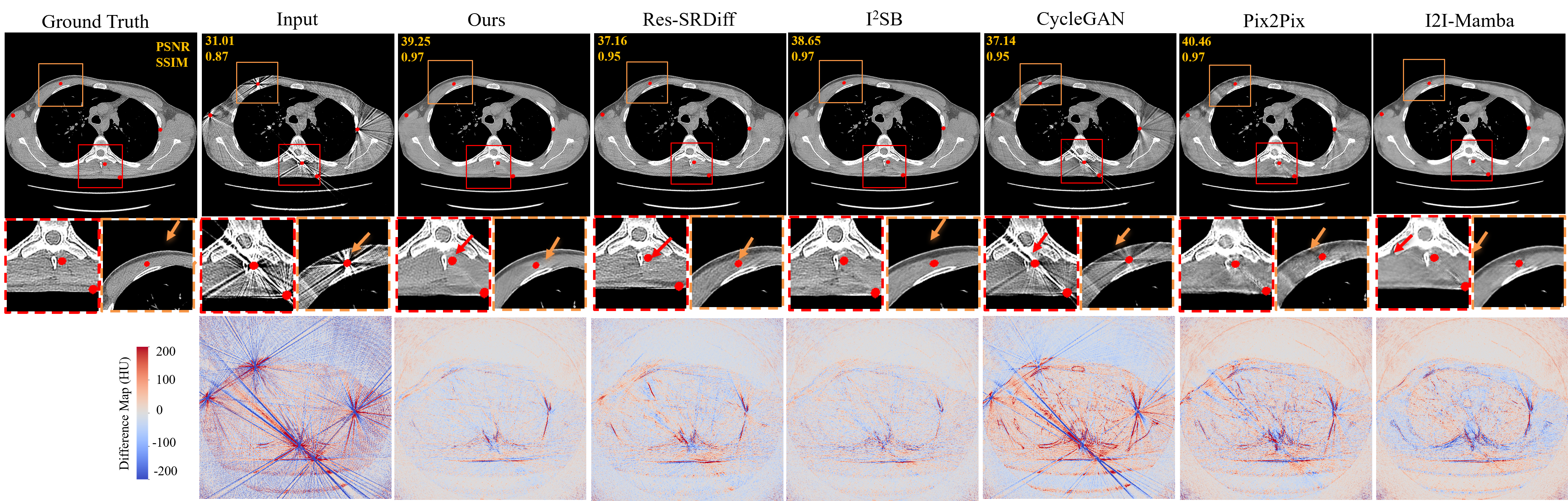}
    \caption{Qualitative comparison on representative thoracic CT slice 261
    from the held-out test set. Reconstructed images and corresponding
    difference maps relative to the metal-free reference are shown for the
    evaluated methods. Red regions indicate metallic implants. CT images are
    displayed using a window of $[-300,300]$ HU, and difference maps are
    reported in HU.}
    \label{fig:result2}
\end{figure}

Figures~\ref{fig:result1} and~\ref{fig:result2} show representative artifact reduction results from the head and thoracic regions. The reconstructed images are compared with the corresponding metal-free references, and the difference maps show the remaining HU deviations after artifact reduction compared to the reference.

Figure~\ref{fig:result1} presents the representative head CT example. The metal-artifact input contains pronounced streak artifacts extending from the metallic regions into the surrounding anatomy. All evaluated methods reduced the dominant artifacts to varying degrees. CycleGAN and Pix2Pix retained larger residual intensity deviations, whereas Res-SRDiff and I2I-Mamba provided stronger  artifact suppression. I$^2$SB produced a reconstruction closely matching the metal-free reference. The proposed diffusion bridge model further reduced residual intensity errors while preserving local anatomical structures. The corresponding difference map showed limited HU deviations in the artifact-affected regions.

Figure~\ref{fig:result2} presents the representative thoracic CT example. The metal-artifact input contains streak artifacts extending from the metallic regions across the surrounding anatomy. CycleGAN and Pix2Pix reduced the major streak patterns but retained visible local intensity differences. Res-SRDiff, I2I-Mamba, and I$^2$SB showed stronger artifact reduction and closer agreement with the metal-free reference. The proposed model effectively reduced the streak artifacts while preserving local anatomical structures and soft-tissue contrast. Its difference map also showed reduced residual HU deviations in the artifact-affected regions.

\subsection{Quantitative and Statistical Comparison}

Table~\ref{tab:quantitative_results} summarizes the quantitative performance of the evaluated methods on the 1,322 held-out test slices. The original metal-artifact images had a PSNR of $29.48\pm3.59$ dB, an SSIM of $0.7972\pm0.1232$, an MAE of $57.33\pm22.14$ HU, and an LPIPS of $0.2912\pm0.1401$.

Among the reconstruction methods, our proposed model achieved the highest PSNR of $43.67\pm4.56$ dB, the highest SSIM of $0.9856\pm0.0138$, and the lowest MAE of $14.74\pm8.68$ HU. I$^2$SB provided the closest performance for these three metrics, with a PSNR of $43.48\pm4.35$ dB, an SSIM of $0.9837\pm0.0144$, and an MAE of $16.59\pm9.87$ HU. I2I-Mamba achieved a PSNR of $40.41\pm3.37$ dB, an SSIM of $0.9752\pm0.0163$, and an MAE of $20.57\pm9.48$ HU.

The ranking differed for LPIPS. Res-SRDiff achieved the lowest LPIPS ($0.0150\pm0.0121$), followed by I$^2$SB ($0.0169\pm0.0186$), whereas the proposed model obtained an LPIPS of $0.0401\pm0.0564$. These results indicate that the proposed model achieved the strongest performance in image fidelity, structural similarity, and HU accuracy, while Res-SRDiff achieved the lowest perceptual feature-space error.

\begin{table*}[t]
\centering
\caption{Quantitative comparison on the held-out CT-MAR test set. Values are reported as mean $\pm$ standard deviation. SSIM and LPIPS are displayed in units of $10^{-2}$. Bold values indicate the best mean performance for each metric. Friedman tests compare the six reconstruction methods, excluding the artifact input; $^{***}p<0.001$ indicates an overall difference among methods, not pairwise significance. Kendall's $W$ quantifies the omnibus effect size.}
\label{tab:quantitative_results}
\resizebox{\textwidth}{!}{%
\begin{tabular}{lcccc}
\toprule
Method &
PSNR (dB) $\uparrow$ &
SSIM ($\times10^{2}$) $\uparrow$ &
MAE (HU) $\downarrow$ &
LPIPS ($\times10^{2}$) $\downarrow$ \\
\midrule
Input (artifact) &
29.48 $\pm$ 3.59 &
79.72 $\pm$ 12.32 &
57.33 $\pm$ 22.14 &
29.12 $\pm$ 14.01 \\
\midrule
CycleGAN &
34.32 $\pm$ 3.72 &
93.02 $\pm$ 4.59 &
36.32 $\pm$ 15.24 &
8.84 $\pm$ 5.68 \\

Pix2Pix &
38.54 $\pm$ 3.56 &
96.25 $\pm$ 2.16 &
24.19 $\pm$ 10.40 &
6.03 $\pm$ 3.16 \\

Res-SRDiff &
39.89 $\pm$ 3.22 &
95.00 $\pm$ 2.31 &
23.15 $\pm$ 9.07 &
\textbf{1.50 $\pm$ 1.21} \\

I2I-Mamba &
40.41 $\pm$ 3.37 &
97.52 $\pm$ 1.63 &
20.57 $\pm$ 9.48 &
5.13 $\pm$ 4.37 \\

I$^2$SB &
43.48 $\pm$ 4.35 &
98.37 $\pm$ 1.44 &
16.59 $\pm$ 9.87 &
1.69 $\pm$ 1.86 \\

RDBM-MAR (ours) &
\textbf{43.67 $\pm$ 4.56} &
\textbf{98.56 $\pm$ 1.38} &
\textbf{14.74 $\pm$ 8.68} &
4.01 $\pm$ 5.64 \\
\midrule
Friedman $\chi^2(5)$ &
$5818.91^{***}$ &
$5650.74^{***}$ &
$5980.92^{***}$ &
$4779.70^{***}$ \\

Kendall's $W$ &
0.880 &
0.855 &
0.905 &
0.723 \\
\bottomrule
\end{tabular}
}
\end{table*}

\begin{figure}[t]
    \centering
        \includegraphics[
            width=\textwidth,
            keepaspectratio
        ]{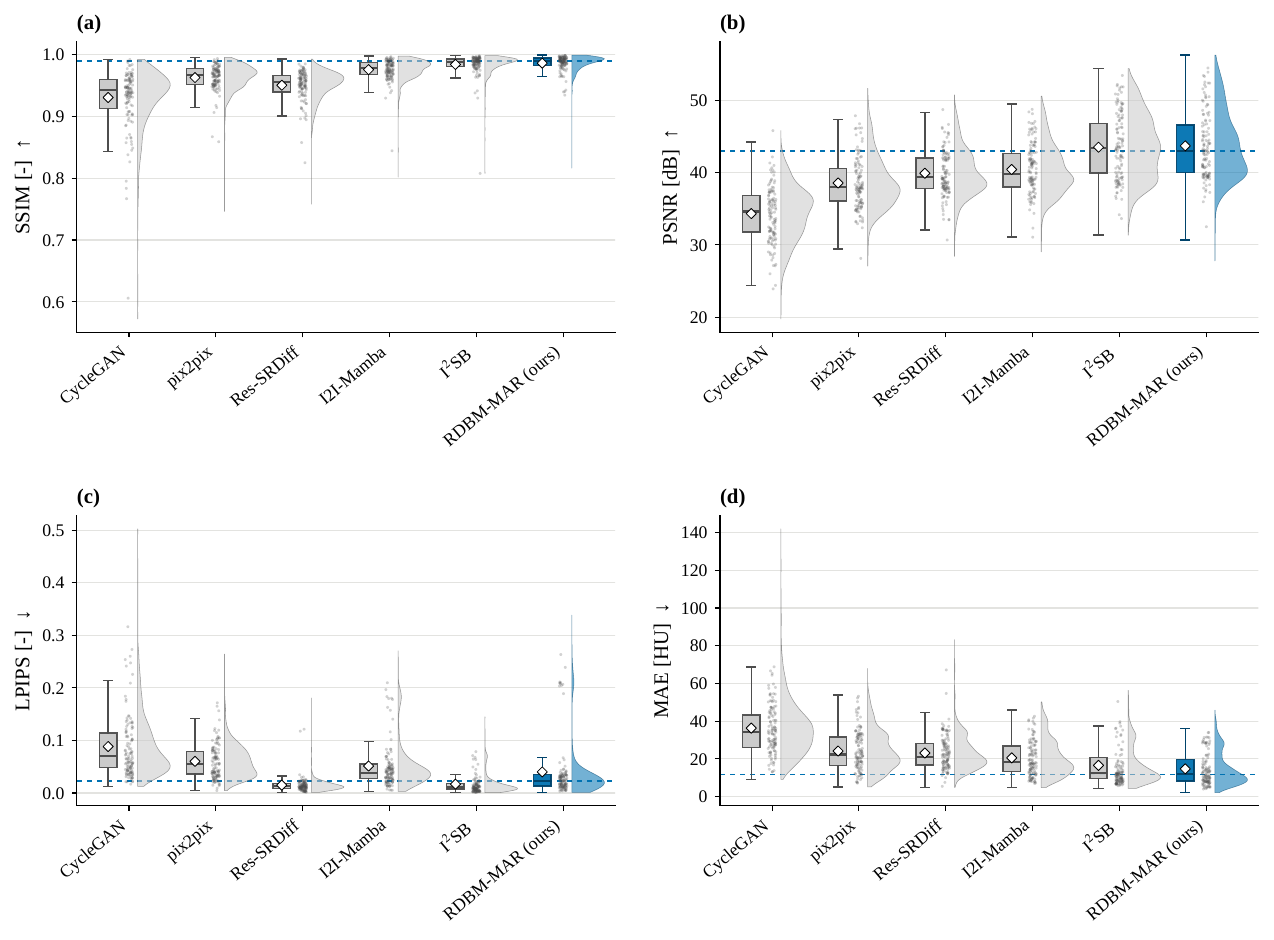}
    \caption{Figure 4: Slice-level distributions of (a) SSIM, (b) PSNR, (c) LPIPS, and (d) MAE across the 1,322 held-out test slices. Higher values indicate better performance for PSNR and SSIM, whereas lower values indicate better
    performance for MAE and LPIPS.}
    \label{fig:boxplots}
\end{figure}

Figure~\ref{fig:boxplots} shows the slice-level distributions of PSNR, SSIM, MAE, and LPIPS across the held-out test set. The proposed model maintained favorable PSNR, SSIM, and MAE distributions across the test slices, while substantial overlap with the strongest comparison methods remained for some metrics. The LPIPS distributions followed a different pattern, with Res-SRDiff and I$^2$SB shifted toward lower values. The slice-level variability and overlap among methods motivated the paired statistical analysis rather than relying solely on the mean values reported in Table~\ref{tab:quantitative_results}.

The Friedman omnibus test showed significant differences among the six reconstruction methods for all four image-quality metrics. Significant differences were observed for PSNR ($\chi^2(5)=5818.91$, $p<0.001$), SSIM ($\chi^2(5)=5650.74$, $p<0.001$), MAE ($\chi^2(5)=5980.92$, $p<0.001$), and LPIPS ($\chi^2(5)=4779.70$, $p<0.001$). Kendall's $W$ values were 0.880, 0.855, 0.905, and 0.723 for PSNR, SSIM, MAE, and LPIPS, respectively, indicating substantial between-method differences across the paired test slices.

Pairwise Wilcoxon signed-rank tests with Holm correction showed that the proposed model significantly outperformed each of the five comparison methods in PSNR, SSIM, and MAE (all adjusted $p<0.001$). Compared with I$^2$SB, the closest-performing comparison method for these metrics, the proposed model achieved median paired differences of $+0.556$ dB in PSNR, $+0.0014$ in SSIM, and $-1.557$ HU in MAE. For LPIPS, the proposed model achieved significantly lower values than CycleGAN, Pix2Pix, and I2I-Mamba, but significantly higher values than Res-SRDiff and I$^2$SB after Holm correction.

\subsubsection{HU Profile Analysis}

\begin{figure}[t]
    \centering
        \includegraphics[width=1\textwidth]{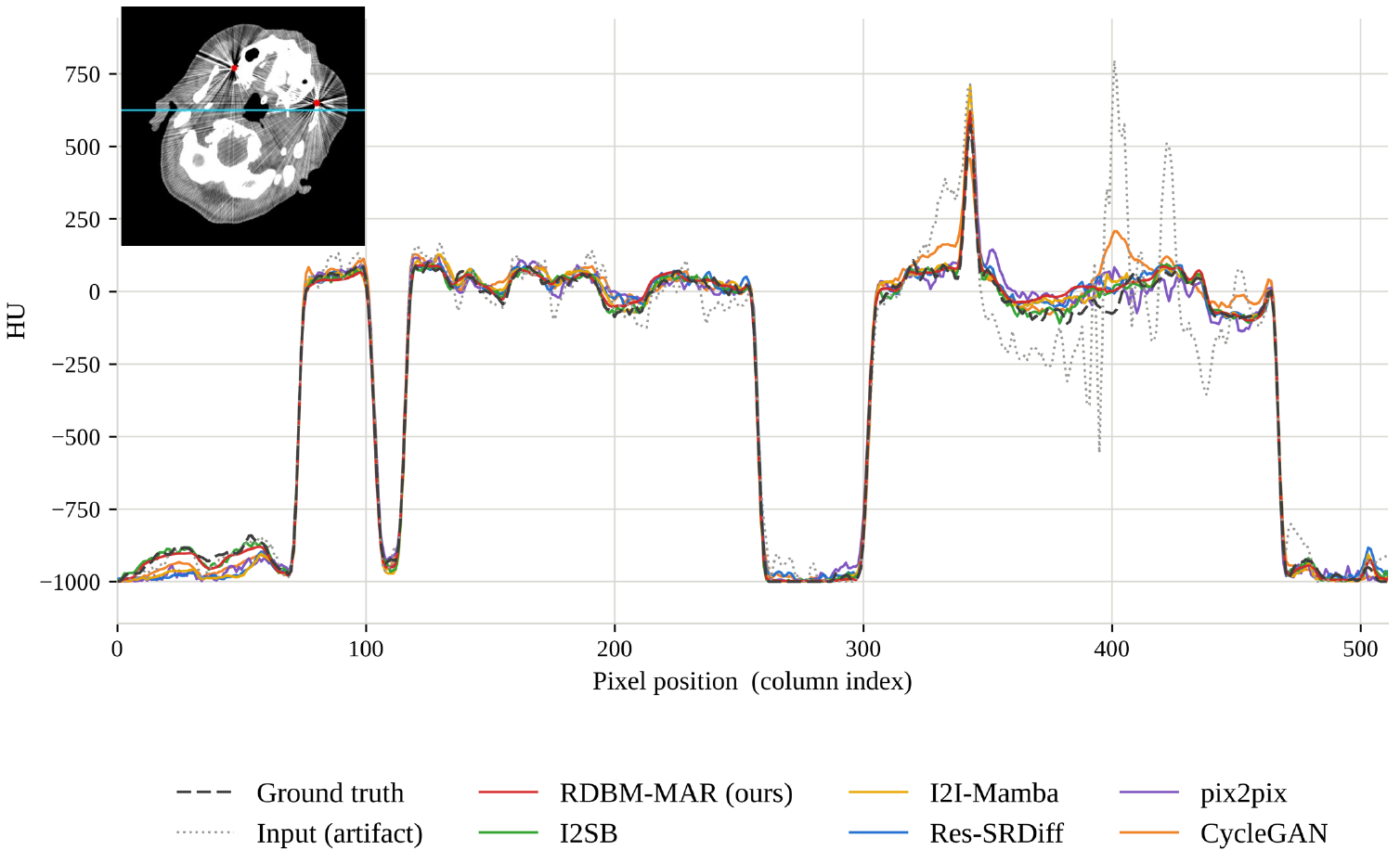}
    \caption{HU profile comparison for representative head CT. Intensity profiles from the metal-artifact input  and reconstructed images are compared with the corresponding metal-free reference along the selected line.}
    \label{fig:result3}
\end{figure}

Figure~\ref{fig:result3} provides an HU profile comparison for the same head CT example shown in Fig.~\ref{fig:result1}. The metal-artifact input showed substantial deviations from the metal-free reference along this profile. Although all reconstruction methods reduced these deviations to varying degrees, differences remained around the regions with the strongest intensity variation. The profile produced by the proposed model more closely followed the reference CT values across the artifact-affected region. This local profile analysis complements the whole-slice MAE by illustrating how the reconstruction behaves across individual anatomical and artifact boundaries.

\subsection{Clinical CT Evaluation}

\begin{figure}[!tbh]
    \centering
        \includegraphics[
            width=\textwidth,
        ]{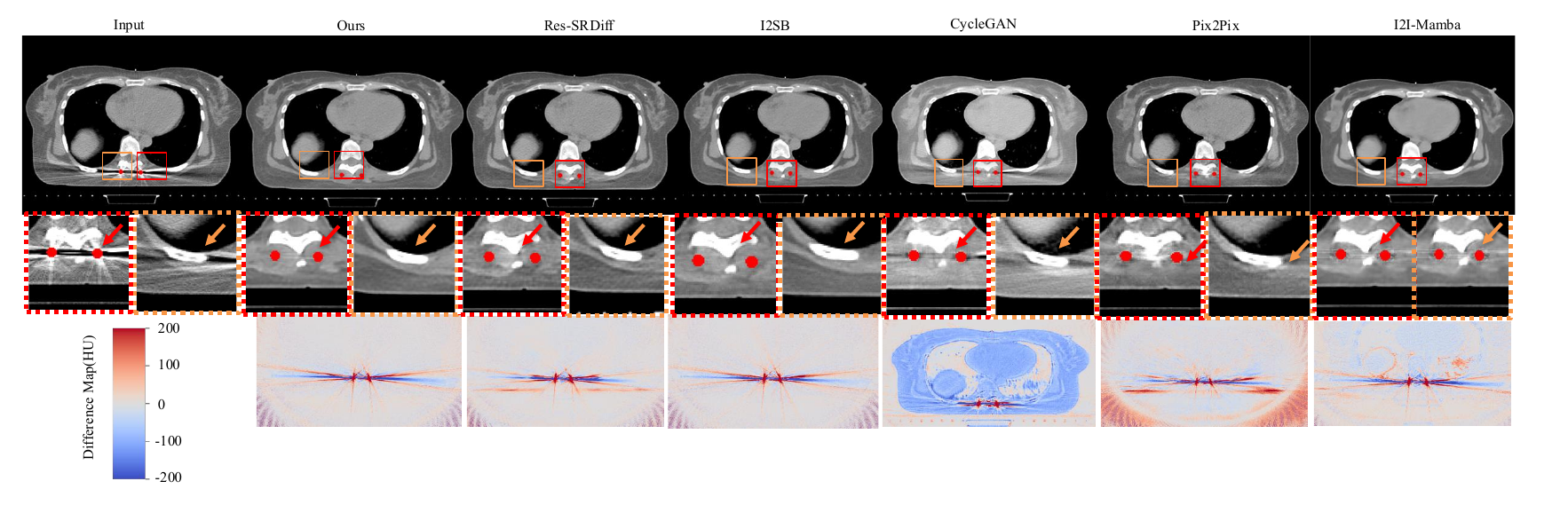}
    
    \caption{Qualitative evaluation on clinical image~13 from the AAPM
    CT-MAR scoring dataset. Metal regions identified from the original
    CT image using a threshold of 3000 HU are overlaid in red for
    visualization. CT images are displayed using a window of
    $[-300,300]$ HU. No metal-free reference is available for this clinical case. Difference maps were calculated as the reconstructed image minus the input metal-artifact CT image.}
    \label{fig:result4}
\end{figure}

Figure~\ref{fig:result4} shows the application of the proposed model to clinical images from the AAPM CT-MAR scoring dataset.  Because a metal-free reference image is not available for this clinical case, the evaluation was conducted qualitatively. The input contains visible metal-associated streak artifacts surrounding the implant regions. After reconstruction, the proposed model reduced the prominent streaking and improved the visibility of the surrounding anatomical structures.

Metal regions identified from the original CT image using a threshold of 3000 HU are displayed as red overlays for visualization only. The metal mask was not provided to the model during training or inference. This clinical example provides an additional qualitative evaluation on clinically representative CT data containing real metal artifacts.

\section{Discussion}

The proposed diffusion bridge model demonstrated strong performance for image-domain CT metal artifact reduction across the held-out test set. Among the six reconstruction methods, the proposed model achieved the highest PSNR and SSIM and the lowest MAE, indicating closer agreement with the metal-free reference images in terms of  overall image fidelity, structural similarity, and CT intensity accuracy. The paired statistical analysis further supported these findings. The Friedman tests showed significant differences among the evaluated methods for all four metrics, with Kendall's $W$ values ranging from 0.723 to 0.905. Pairwise Wilcoxon signed-rank tests with Holm correction showed that the proposed model significantly outperformed all five comparison methods in PSNR, SSIM, and MAE. These results indicate that the observed improvements were consistent across the paired test slices rather than being driven only by differences in average performance.

The qualitative results were consistent with the quantitative evaluation. In both the representative head and thoracic CT examples, the proposed model substantially reduced metal-induced streak artifacts while preserving the surrounding anatomical structures. The corresponding difference maps showed smaller residual HU deviations in artifact-affected regions. The HU line-profile analysis further showed that the reconstructed CT values preserved the overall intensity trend of the metal-free reference with reduced local fluctuations. Together with the lower MAE observed across the full test set, these findings support the quantitative reliability of HU recovery by the proposed model and suggest that artifact-related intensity errors can be reduced with lower risks of altering the underlying anatomical information.

The ranking of the methods differed for LPIPS. Res-SRDiff achieved the lowest LPIPS, followed by I$^2$SB, whereas the proposed model achieved the best performance in PSNR, SSIM, and MAE. Pairwise testing similarly showed that the proposed model had lower LPIPS than CycleGAN, Pix2Pix, and I2I-Mamba, but higher LPIPS than Res-SRDiff and I$^2$SB. This difference indicates that the similarity of the learned perceptual features and the accuracy of CT intensity do not necessarily rank the reconstruction methods in the same order, consistent with the perception--distortion trade-off in image restoration~\cite{blau2018tradeoff}. In the reduction of CT metal artifacts, this distinction is important because accurate recovery of HU values and anatomical structures is required in addition to visual similarity. The lower MAE of the proposed method therefore provides complementary information to the LPIPS results. Unlike natural-image restoration, CT reconstruction requires preservation not only of visual appearance but also of quantitatively meaningful image intensities. HU values carry crucial physical information that is used in quantitative image interpretation and clinical analysis. In this context, the lower MAE achieved by the proposed model indicates improved recovery of CT intensity values and complements the information provided by perceptual metrics such as LPIPS. The simultaneous improvement in PSNR, SSIM, and MAE therefore suggests that the proposed method reduces artifact-related errors while maintaining both structural and quantitative fidelity to the metal-free reference.

The observed performance may be related to the formulation of the restoration process. Unlike conventional diffusion approaches that initialize reverse reconstruction from an independent Gaussian sample, the diffusion bridge starts directly from the observed metal-artifact CT image. As a result, anatomical information that remains present in the degraded observation is retained at the beginning of reconstruction rather than being regenerated from noise. The conditional U-Net further predicts an image-domain correction relative to the metal-artifact image, focusing the reconstruction on the differences between the degraded observation and the metal-free target. The bridge formulation therefore provides a direct restoration trajectory between the paired artifact and metal-free endpoints. In addition, the shortened deterministic reverse trajectory allows reconstruction to be performed with a limited number of network evaluations. Although the contribution of each component was not separately evaluated in the current study, these design characteristics may contribute to the strong intensity and structural fidelity observed in the results.

The clinical CT example from the AAPM CT-MAR scoring dataset provided an additional qualitative evaluation beyond the paired benchmark data. The model reduced visible streak artifacts surrounding the metal regions and improved the visibility of nearby anatomical structures. However, a metal-free reference was not available for this case, and the clinical result should therefore be interpreted qualitatively rather than as a quantitative validation of performance on real metal artifacts.

Several limitations remain. The quantitative evaluation was performed using paired benchmark data generated through the AAPM CT-MAR simulation framework, and performance on clinical CT images may differ because of variations in scanner configuration, implant composition, anatomy, and artifact severity. The current model also processes each CT slice independently and does not use information from adjacent slices or three-dimensional anatomical continuity. In addition, the clinical evaluation was limited to a single scoring case without a metal-free reference. Future work will therefore include evaluation on a larger range of clinical CT images and implant types, reader-based assessment, and extension of the framework to adjacent-slice or three-dimensional reconstruction. Metal masks and projection-domain measurements could also be incorporated as additional conditioning information. Finally, alternative training objectives may help improve perceptual similarity while preserving the HU accuracy and structural fidelity achieved by the current model.

\section*{Conflicts of interest}
There are no conflicts of interest declared by the authors.

\section*{Acknowledgment}

This research is supported in part by the National Institutes of Health under Award Numbers R01HL183223, R01DE033512, and R01CA272991.

This work used the datasets provided by the AAPM CT-MAR Grand Challenge. The training datasets were generated using a hybrid data simulation framework, which incorporates publicly available clinical
images from the DeepLesion CT dataset and the UCLH Stroke EIT Dataset, together with metal objects generated using the CatSim simulator in the open-source toolkit XCIST. This work also used clinical CT data from the benchmark provided by the AAPM CT-MAR Grand Challenge. The dataset and associated benchmark
resources are available through the official challenge website:
\url{https://www.aapm.org/GrandChallenge/CT-MAR/Default.asp}.

\bibliographystyle{spiebib}
\bibliography{reference_moj_v02.bib}

\end{document}